# Spin polarization driven by molecular vibrations leads to enantioselectivity in chiral molecules

Shinji Miwa[1,2*], Tatsuya Yamamoto[3], Takashi Nagata[1], Shoya Sakamoto[1], Kenta Kimura[4], Masanobu Shiga[1], Weiguang Gao[1], Hiroshi M. Yamamoto[5], Keiichi Inoue[1], Taishi Takenobu[6], Takayuki Nozaki[3] & Tatsuhiko Ohto[6**]

1. The Institute for Solid State Physics, The University of Tokyo, Kashiwa, Chiba 277-8581, Japan
2. Trans-scale Quantum Science Institute, The University of Tokyo, Bunkyo, Tokyo 113-0033, Japan
3. National Institute of Advanced Industrial Science and Technology (AIST), Research Center for Emerging Computing Technologies, Tsukuba, Ibaraki 305-8568, Japan
4. Department of Materials Science, Osaka Metropolitan University, Osaka 599-8531, Japan
5. Institute for Molecular Science, Okazaki, Aichi 444-8585, Japan
6. Graduate School of Engineering, Nagoya University, Nagoya, Aichi 464-8603 Japan
[*]miwa@issp.u-tokyo.ac.jp
[**]ohto@nagoya-u.jp

**Chirality pervades multiple scientific domains−physics, chemistry, biology, and astronomy−and profoundly influences their foundational principles[1,2]. Recently, the chirality-induced spin selectivity (CISS) phenomenon[3-5] has captured significant attention in physical chemistry due to its potential applications and intriguing underlying physics[3-11]. Despite its prominence, the microscopic mechanisms of CISS remain hotly debated[12-15], hindering practical applications and further theoretical advancements. Here we challenge the established view that attributes CISS-related phenomena to current-induced spin polarization and electron transport across interfaces[3,5,12-14]. We propose that molecular vibrations in chiral molecules primarily drive spin polarization, thereby governing CISS. Employing an electrochemical cell paired with a precisely engineered magnetic multilayer, we demonstrate that the magnetic interactions akin to interlayer exchange coupling[16-19] are crucial for CISS. Our theoretical study suggests that molecular vibrations facilitate chirality-dependent spin polarization, which plays a pivotal role in CISS-related phenomena such as magnetoresistance[8] and enantiomer separation[9] using ferromagnets. These findings necessitate a paradigm shift in the design and analysis of systems in various scientific fields, extending the role of spin dynamics from traditional areas such as solid-state physics to chemical reactions, molecular biology, and even drug discovery.**

**Introduction**

Chirality, defined as a pseudo-scalar that changes signs under mirror reflection, is characterized by the lack of mirror symmetry[20]. This intriguing property is widely recognized and studied in diverse fields such as physics, chemistry, biology, and astronomy[1,2]. The impact of chirality on chemical reaction yields is a well-established concept, and the origins and roles of homochirality in biomolecules have sparked extensive debate and research. Recently, numerous phenomena related to chirality-induced spin selectivity (CISS) have been reported in the field of physical chemistry[3-5]. One seminal observation in this area was the detection of photoelectrons that, upon passing through double-stranded DNA molecules, exhibited finite spin angular momentum in a Mott polarimeter[6,7]. This discovery marked the beginning of a series of studies and reports on various CISS-related phenomena. Particularly noteworthy among these phenomena are the manifestation of magnetoresistance in junctions involving chiral molecules and ferromagnetic electrodes[8], as well as the enantiomer separation using ferromagnetic substrates[9]. These findings are not only intriguing from a scientific standpoint but also hold significant potential in practical applications, especially in the realm of enantioselective synthesis.

Longitudinal spin current, which represents a flow of spin angular momentum parallel or antiparallel to the current direction, exemplifies a truly chiral system[21]. In the context of CISS-related phenomena, it has been considered that electrons traversing chiral materials gain orbital angular momentum due to their helical motion, subsequently achieving spin polarization via spin–orbit interaction. This process is similar to magnetochiral anisotropy[22]. However, the model explaining CISS through current-induced spin polarization presents certain shortcomings. Within the linear response domain, the occurrence of magnetoresistance is theoretically prohibited[14], and these models also contravene Onsager's reciprocity principle, as evidenced by symmetry-based arguments[15]. Nevertheless, experimental studies have reported exceptionally very large magnetoresistance[23,24], surpassing those observed in commercially available spintronic devices, such as the CoFeB/MgO system[25-27]. Concerning the enantiomer separation facilitated by achiral ferromagnets, a model has been suggested where transient currents generated during molecular adsorption on metals induce current-induced antiparallel pair of spin polarizations in chiral molecules[9]. This spin polarization is believed to persist over extended periods[28]. Yet, in practice, nonequilibrium spin polarization in materials typically exhibits very short spin relaxation times[29]. Consequently, a comprehensive and unified understanding of CISS-related phenomena remains elusive, posing a significant challenge to its practical application and further advancement in the field.

Here we employed the prototypical chiral electrolyte, (1$S$)-(+)- or (1$R$)-(−)-camphor-10-sulfonic acid[30] (($S$)- or ($R$)-CSA, Fig. 1**a**), alongside a precisely engineered ferromagnetic CoPt/Au electrodes, to conduct time-resolved observation of magnetoresistance. We find that the essence of CISS lies in the magnetic interaction between chiral molecules and the ferromagnetic electrode, analogous to the interlayer exchange coupling[19]. This is mediated by conduction electrons through the Ruderman-Kittel-Kasuya-Yosida (RKKY) interaction[16-18]. Our findings reveal a critical insight: in CISS-related phenomena, the electric current is not used to polarize the spins, but is merely a probe of the system. The spin polarization driven by molecular vibration of chiral molecules plays a pivotal role (Fig. 1**a**). This vibration-driven spin polarization aligns with signature of thermally driven spin polarization[31-33] and observations of CISS-related phenomena occurring without a bias current in the system[32-36]. Furthermore, the concept is consistent with the enhancement of CISS with increasing temperature[37]. While previous studies have noted that molecular vibration can enhance current-induced spin polarization[38-40], our work posits that molecular vibration can create magnetic interaction in the system.

**Experimental design and typical characteristics of the electrochemical cell**

Electric measurements were carried out using a custom-made electrochemical cell (refer to Methods and Extended Data Fig. 1 for details). To measure the magnetoconductance (MC) effect attributable to CISS, it is crucial to align the magnetization direction of the working electrode with the direction of the electric current flow, essentially perpendicular to the working electrode's film plane. CoPt, a ferromagnetic material known for its perpendicular magnetization (Fig. 1**b**), was selected for its notable resistance to corrosion in electrolyte environments. This characteristic significantly favors CoPt over commonly used Ni in CISS-related studies. To protect the CoPt surface and to address the interaction between CoPt and the electrolyte, a Au spacer layer was implemented, constructing CoPt/Au working electrode. Figure 1**b** shows magnetization hysteresis curves of the CoPt/Au(2 nm) electrode, obtained through magneto-optical Kerr effect measurements, both before and after the evaluation of the MC effect. These curves reveal no alteration in the coercive field of CoPt following electrical measurement in an electrolyte solution with 0.25 mM CSA, confirming that the CoPt/Au has a resistance to corrosion in an electrolyte environment. Figure 1**c** depicts the setup for MC measurement using the electrochemical cell, utilizing the CoPt/Au electrode as the working electrode. The electrolyte solution was prepared with chiral electrolyte CSA and supporting electrolyte KCl. By employing a CSA concentration of 0.25 mM, which is significantly lower than that employed in prior studies[30], our experimental design aimed to minimize the corrosive effects of the electrolyte on the CoPt/Au electrode and ensure a well-defined interface between the CoPt/Au electrode and CSA.

Figure 1**d** illustrates the typical current–voltage characteristics observed in our electrochemical cell. When a voltage was applied in the negative direction starting from 0 V, the electric current exhibited a peak at −1.4 V. In this configuration, under the applied negative voltage, the working electrode functions as an anode, facilitating the movement of electrons from the working electrode to the electrolyte solution. This anode reaction is crucial in determining the overall electric current in the system, primarily due to the working electrode's surface area being designed to be significantly smaller than that of the counter electrode, as shown in Extended Data Figs. 1**a** and 1**b**. The current in the electrochemical cell significantly increases with the addition of CSA (see Extended Data Figs. 1**c** and 1**d**), hence the electrode reaction characterized by the current peak around −1.4 V is due to the oxidation (reduction) in CoPt/Au (CSA). Since the reduction involves CSA accepting electrons to facilitate the conversion of camphor to borneol, this electrode reaction is considered dominant in the results of Fig. 1**d**. For the evaluation of MC effect, which is the variation in current as a function of the magnetization direction of CoPt, the voltage was set at $V_{meas}$. The electric current was then measured under this constant voltage setting (chronoamperometry). It is noteworthy that $V_{meas}$ exhibits a dependence on the thickness of the Au film, showing a slight increase as the thickness of the Au increases (Extended Data Fig. 1**e**).

Upon applying a constant voltage ($V_{meas}$) to the electrochemical cell, an initial surge of relatively large current (>1 μA) is observed, which then exhibits an abrupt decline as shown in Fig. 1**d** inset. This phenomenon can be attributed to two primary factors. The first is the reduction in the amount of ion movement in the solvent. As a constant voltage is applied, ions move in the electrolyte solution, forming an electric double layer near the electrode surface. As this double layer forms, the movement of ions decreases, leading to a reduction in current. The second reason is the decrease in the electrolyte concentration near the electrode surface due to electrochemical reactions. The electrolyte at the electrode surface is used by the electrochemical reactions. When this consumption exceeds the supply of the electrolyte from the bulk solution through diffusion, the concentration of electrolyte near the electrode decreases, resulting in a decrease in current. Over time, as these processes reach a balance, the current stabilizes, primarily constrained by the diffusion rate of the electrolyte to the electrode surface. In this quasi-steady state, the presence of 0.25 mM CSA significantly influences the current, with more than a twofold difference observed (Extended Data Fig. 1**d**). This variation underscores that the electric current is predominantly driven by the electrochemical reaction at the CoPt/Au electrode interface with CSA. The MC effect was assessed in this quasi-stable condition.

Figure 1**e** shows typical results demonstrating the MC effect. A constant magnetic field of 0.6 T was applied perpendicular to the electrode surface using an electromagnet. The field's polarity was alternated every 300 seconds, with each transition taking approximately 5 seconds. The data show that switching magnetic field's polarity from negative to positive (and vice versa) leads to a gradual decrease (increase) in current until a new steady state is established, with a relaxation time of about 50 seconds. Here, the MC effect diverges from predictions made by traditional theories, such as spin-dependent tunneling, akin to the tunnel magnetoresistance effect[25-27]. If changes in interface conductance due to spin-dependent tunneling were responsible, we would expect an immediate response in current following a change in the magnetic field's polarity, with the relaxation trends counteracting this change. Contrarily, our experimental results indicate a monotonous decrease or increase in conductance following the change in magnetic field's polarity, suggesting that a mechanism other than spin-dependent tunneling conduction is required to explain this MC effect. In fact, the electric current in the quasi-steady state is governed by physical parameters related to the bulk electrolyte concentration and diffusion coefficient[41]. Given the unlikely dependence of bulk electrolyte concentration on magnetization direction, we propose that the diffusion coefficient of CSA in proximity to the CoPt/Au interface is influenced by the direction of magnetization. This variation alters the CSA supply rate to the electrode surface, thereby affecting the local CSA concentration and, consequently, the electric current. A deeper analysis of this phenomenon and its implication will be provided in subsequent discussions.

**Time-resolved magnetoconductance (MC) measurements**

To elucidate the mechanism behind the MC effect, Figure 2 explores the chirality dependence of CSA, the impact of working electrode size, and the effect of the thickness of the Au spacer layer. Figure 2**a** details the changes in conductance over 300 seconds following a switch in the magnetic field's polarity, employing CoPt/Au(1 nm) electrode setup akin to the methodology described in Fig. 1**e**. The MC effect was defined by $(I_{B = +0.6\ T} - I_{B = -0.6\ T})/(I_{B = +0.6\ T} + I_{B = -0.6\ T})$. Notably, the experiment reveals that (*S*)-CSA induces a positive MC effect, whereas (*R*)-CSA resulted in a contrasting negative MC effect. In trials using a racemic mixture of CSA, the MC effect was observed to be minimal. These findings align with the CISS-related phenomena reported in prior studies, underscoring the critical role of molecular chirality in determining the polarity of the MC effect.

Figure 2**b** delves into how the MC effect varies with the surface area of the working electrode. Notably, the MC effect was minimal (< 0.1%) for a small electrode with an area of ~0.1 mm$^2$. However, as the electrode size increased, a significant MC effect (> 1% at 0.5 mm$^2$) was observed and it was subsequently used for measurements other than Fig. 2**b**. The observed tendency of the MC effect on the electrode area challenges the notion that its origin lies in spin-dependent tunneling conduction, as the MC ratio would not be expected to vary with electrode size under such mechanism. The key factor influencing the rate of electrode reaction with changes in electrode size is the alternation in electrolyte diffusion. For larger electrodes, significantly exceeding the thickness of diffusion layer (Gouy-Chapman layer), i.e. the area near the electrode where electrolyte concentration diverges from the bulk concentration, electrolyte diffusion tends towards a one-dimensional model, perpendicular to the electrode surface. Conversely, for smaller electrodes not meeting this criterion, the edge effect becomes significant, leading to three-dimensional diffusion and an enhanced rate of electrolyte supply to the electrode interface. This increased supply with the smaller electrode size could explain

the observed reduction in the MC effect. Therefore, the experiment posits that the MC effect originates from interactions tied to the magnetization direction of CoPt/Au and the chirality of CSA, influencing the CSA concentration near the electrode surface. This alteration in concentration precedes the electrochemical reaction, i.e., the electron transfer process from the CoPt/Au electrode to CSA, suggesting the presence of a magnetic interaction between CoPt/Au and CSA independent of electric current flow. Unraveling the nature and origin of this magnetic interaction is pivotal for a comprehensive understanding of the MC effect.

To elucidate the nature of the magnetic interaction underlying the MC effect, we explored how the MC effect varies with the thickness of the Au spacer layer. Figure 2**c** details the characteristics of the MC effect's dependence on Au thickness, with the corresponding MC ratio depicted in Fig. 2**d**. Notably, Fig. 2**d** reveals that the MC ratio starts off negative at nearly zero Au thickness and exhibits oscillation with a period of ~2 nm, including changes in the sign. This oscillatory behavior is analogous to the interlayer exchange coupling[19] observed in Fe/Au/Fe systems, which has been reported to have a periodicity of ~1.8 nm[42,43]. The observation from Fig. 2**d** suggest the presence of RKKY oscillation[16-18] mediated by conduction electrons within the Au spacer layer, facilitating magnetic interaction between CoPt and CSA. This leads to the conclusion that the chirality-dependent spin polarization of CSA, independent of electric current flow, plays a pivotal role in the manifestation of the MC effect. As indicated above, the MC effect observed in Fig. 2**b** is characteristic at a CSA concentration of 0.25 mM. This effect remains almost unchanged even with increased CSA concentrations; a comparable MC effect was noted at concentrations up to 4 mM CSA (refer to Extended Data Figs. 2**a** and 2**b**).

In the measurements of the time-resolved MC effect, we first saturate the magnetization of the CoPt electrodes to positive (negative) by applying a magnetic field of +0.6 T (−0.6 T) and then maintain this magnetic field strength during the measurements. On the other hand, CoPt is perpendicularly magnetized, and the magnetization at zero magnetic field after saturation at ±0.6 T is almost the same as that at ±0.6 T. Therefore, if the effect of the external magnetic field was solely to change the direction of magnetization of the CoPt electrodes, the MC effect should be observed regardless of whether the magnetic field is applied during the measurements. Interestingly, when we saturate the magnetization of CoPt with a magnetic field of ±0.6 T and then conduct the measurement at zero magnetic field, no MC effect is observed (see Extended Data Fig. 2**b**). Thus, the absence of the MC effect in the zero magnetic field measurement strongly suggests that the external magnetic field contributes not only to the magnetization reversal in CoPt but also to the induction of chirality-dependent spin polarization in molecules.

The MC ratio depicted in Fig. 2 peaks at ~3%, which is lower than the ~20% MC effect reported in previous studies utilizing cyclic voltammetry measurements with Ni/Au(10 nm) electrodes and a higher concentration of CSA (20 mM)[30]. This discrepancy prompted an investigation into how the CSA concentration affects the MC effect. For example, employing a 0.25 mM CSA solution with a CoPt/Au(1 nm) electrode setup resulted in a +1.1% MC effect as shown in Fig. 2**a**, whereas using an 8 mM CSA solution yielded a −3.2% MC effect, effectively tripling the MC ratio, as demonstrated in Extended Data Figs. 2**c** and 2**d**. However, unlike the stable magnetization curves of CoPt observed with CSA concentrations below 4 mM, a significant alteration in the hysteresis curve was detected following measurements at 8 mM CSA (refer to Extended Data Fig. 2**c**). This suggests that a high CSA concentration can induce a direct reaction between CSA and CoPt, leading to the dissolution of CoPt into the solution despite the presence of an Au layer covering the electrode's surface. Additionally, the shift in the MC effect's polarity from positive at a 0.25 mM CSA concentration to negative at 8 mM CSA in CoPt/Au(1 nm) electrodes indicates a direct electrochemical reaction between CSA and CoPt, unaffected by the Au layer. The relatively modest MC effect in Fig. 2 might arise from the use of a low CSA concentration and the selection of a ferromagnetic electrode (CoPt) known for its corrosion resistance, which helps in maintaining a well-defined metal/solution interface. It appears that using more soluble electrode like Ni or higher electrolyte concentrations tends to enhance the MC effect while compromising the integrity of the interface.

**Theoretical studies on spin polarization in chiral molecules**

To induce the MC effect, it is essential to establish a spin polarization in CSA molecules without an electric current. While previous studies have shown that molecular vibrations can enhance current-induced spin polarization[38-40], they have not demonstrated the generation of spin polarization without an electric current. However, our experiments provide compelling evidence that external magnetic fields significantly contribute to the induction of chirality-dependent spin polarization in molecules. Accordingly, we explored how molecular vibrations under magnetic field influence the spin polarization of chiral molecules, using first-principles calculations (see Methods).

To induce a finite spin density in the CSA molecule, which is a closed shell with no intrinsic spin density, we introduced 0.1 electrons into the molecular structure. The results, depicted in Figs. 3**a** and 3**b**, show the spin density difference isosurfaces with a value of $\pm 0.0001$ $e$ $\text{Å}^{-3}$ for (*S*)-CSA post-energy relaxation, with downward electric polarization and upward (downward) magnetization (spin angular momentum). This figure highlights a molecular vibrational mode at 1770 cm$^{-1}$, characterized by significant vibrational circular dichroism (VCD) and the largest electron–vibration coupling constant (see Extended Data Fig. 3). It shows changes in spin angular momentum (**S**) with positive (+) and negative (−) displacements, yellow for **S** > 0 (upward) and cyan for **S** < 0 (downward). We define displacements as positive (negative) when they lead to a decrease (increase) in the electric polarization of the CSA molecule. With respect to the CSA's carbon and oxygen atoms, which exhibit substantial changes in spin density, these positive (negative) displacements align with clockwise (anticlockwise) movements. Given that molecular vibrations involve alternating displacements in both the + and − directions, the resulting contours can be interpreted as the time derivative of the spin angular momentum (d$\mathbf{S}$/d$t$). Calculations for reversed polarization directions, as well as for the (*R*)-CSA, are presented in Figs. 3**c-h**.

Focusing initially on Figs. 3**a-d**, we notice that the spin densities exhibit opposite polarities in response to + and − displacements. However, given that the half periods, $\tau_+$ and $\tau_-$, are identical, it appears improbable that the CoPt/Au ferromagnetic electrode would magnetically interact with the CSA in a manner influenced by the electrode's spin polarization. This is consistent with the principle that chirality is time-reversal even, and thus, it does not facilitate the induction of spin polarization in an equilibrium state.

It is known that magnetic fields can alter the period of circularly polarized phonons[44,45]. Considering the molecular vibration, we expected differences in the half periods $\tau_+$ and $\tau_-$ due to the magnetic field's influence, which is corroborated by the spin–vibration coupling[45] and vibrational angular momentum. For example, in Figs. 3**a** and 3**b**, we expect $\tau_-$ to be larger than $\tau_+$ and the opposite in Figs. 3**c** and 3**d**. Thus, despite the time integral of d$\mathbf{S}$/d$t$ remaining zero, molecular vibrations introduce a disparity in the durations exhibiting polarities d$\mathbf{S}$/d$t$ > 0 and d$\mathbf{S}$/d$t$ < 0 in (*S*)-CSA, leading to a dominance in the d$\mathbf{S}$/d$t$ density of Figs. 3**a** and 3**d** over 3**b** and 3**c**, as indicated by solid squares in pink. Similarly, for (*R*)-CSA, Figs. 3**f** and 3**g** are more pronounced than 3**e** and 3**h**, as shown by solid squares in blue. The application of a magnetic field thus induces a time-variance in d$\mathbf{S}$/d$t$ polarities based on chirality, which triggers the CISS effect. As CSA molecules approach the CoPt/Au electrode, the electrode responds such that d$\mathbf{S}$/d$t$ < 0 in a positive magnetic field and d$\mathbf{S}$/d$t$ > 0 in a negative one. Therefore, the spin polarizations (d$\mathbf{S}$/d$t$) of both the ferromagnetic electrode and the CSA are influenced by the chirality and the direction of the ferromagnetic electrode's spin polarization, alternating signs based on these interactions and thereby producing a measurable MC effect.

**Discussion**

The concept of spin polarization driven by molecular vibration, as demonstrated in the first-principles study, is schematically represented in Fig. 4. Initially, free-standing chiral molecules, randomly oriented without spin polarization, are shown in Fig. 4a. Upon the application of an electric field (**E**), these molecules align as shown in Fig. 4b. Additionally, applying a magnetic field (**B**) induces a chirality-independent spin polarization (**S**), represented by gray arrows in Fig. 4c. However, this spin polarization does not lead to CISS as it lacks an even function component. At finite temperatures, due to the electron–vibration coupling and spin–orbit interaction, molecular vibrations may induce chirality-dependent spin densities at the edge of the chiral molecules, represented by green arrows in Fig. 4c. These green arrows denote the time derivative of the spin angular momentum (d**S**/d$t$), which changes sign with clockwise (+) and counterclockwise (−) vibrations. Due to the vibrational angular momentum and spin–vibration coupling under an external magnetic field, the half-periods for these vibrations differ, leading to spin polarization polarities (d**S**/d$t$ > 0 or d**S**/d$t$ < 0) driven by molecular vibrations as shown in Fig. 4d. The magnetic field influences both the spin polarization polarity and the relative duration of the half-periods, making the spin polarization independent of the magnetic field direction and dependent solely on molecular handedness.

When comparing CISS-related phenomena with the findings of this study, several key aspects emerge. The spin polarization of photoelectrons traversing chiral molecules[6,7] and the chirality-dependent circular photogalvanic effect[46], attributed to current-induced bulk spin polarization, fall outside the discussion due to symmetry considerations. The use of achiral ferromagnetic metals for enantiomer separation[9] appears to align with the mechanism proposed in this research. Although previous studies have discussed transient current-induced spin polarization during molecule adsorption[9] or under electric fields[4], the extremely short spin relaxation times within materials[29] make this unlikely in practical scenarios. This study proposes that magnetic interactions driven by molecular vibration are significant in achieving enantiomer separation with ferromagnetic electrodes.

Additionally, while previous discussions on magnetoresistance effects at the interface between chiral molecules and ferromagnetic metals focused on current-induced spin polarization in molecules and spin-dependent tunneling at the interface, our study suggests that magnetic interaction driven by molecular vibration could lead to changes in the adsorption state of chiral molecules. These changes might affect the Schottky barrier height at the interface due to structural adjustments, potentially explaining the significant magnetoresistance effects observed, which are not solely attributable to spin-dependent tunneling. Remarkably, studies using identical chiral molecules have reported exceptionally high magnetoresistance ratios (>90%) with conductive atomic force microscope setups, in stark contrast to the substantially lower magnetoresistance effects (<0.1%) observed in multilayer device structures.[23] This discrepancy may be due to the dynamic nature of molecular structures in atomic microscope setups, allowing for easier structural adjustment compared to the more rigid multilayer devices. CISS-induced magnetoresistance effects in studies involving electrolytes[30,47], including this study, tend to show pronounced effects, possibly because the molecules are not rigidly fixed.

The underlying mechanisms of spin polarization in chiral molecules, which are responsible for the bias-current-free CISS phenomena[32-36], have remained elusive. However, spin polarization driven by molecular vibration provides a clear explanation.

**Conclusion**

This study elucidates the mechanism underlying the CISS-induced magnetoresistance effect, pivotal to understanding CISS-related phenomena, via chiral electrolytes. We found that this magnetoresistance originates from spin polarization driven by molecular vibration in chiral molecules and their magnetic interaction with ferromagnetic electrodes. This insight suggest that CISS-related phenomena are more accurately ascribed to this mechanism rather than to current-induced spin polarization, indicating a profound connection not only to CISS but also to broader applications in chemical reactions, molecular biology, and drug discovery. Consequently, a significant shift in system design and analysis across these diverse fields may be necessary, emphasizing the importance of a more comprehensive and integrated approach to exploring and harnessing the intricate relationship between chirality and spin dynamics.

**Methods**

**Custom-made electrochemical cell**

We used a custom-made electrochemical cell, the schematic of which is presented in Extended Fig. 1**a**. The multilayer structure, comprising Ta(5 nm), Pt(10 nm), and [Co(0.25 nm)/Pt(0.15 nm)]$_{20}$ was fabricated on a thermally oxidized silicon substrate using the magnetron sputtering method[48]. After fabrication, the sample was transferred to an electron beam deposition system for further processing. We initiated the process with a soft cleaning of the sample surface by Ar-ion etching, followed by the deposition of Co(0.25 nm) and Au(0-5 nm) layers under ultrahigh vacuum. The multilayer films were then patterned into electrodes suitable for the electrochemical cell, employing conventional photolithography techniques (using resist: AZ6124 and LOR-3A) and Ar-ion etching. As depicted in Extended Data Fig. 1**b**, the working electrodes of various sizes (0.13-0.5 mm$^2$) were prepared, with each working electrode being significantly smaller than the counter electrode (16 mm$^2$). This design ensures that the electric current is limited by the anode reaction, namely, electrochemical reduction (oxidation) of the electrolyte (electrode).

After the microfabrication, the samples underwent a cleaning process. This included an ultrasonic bath with acetone, 2-propanol, and H$_2$O, followed by a 5-minute treatment in a UV cleaner (ASM401N, Asumigiken Ltd.) to remove any residual resist on the sample surface. A custom-designed acrylic wall (Yumoto Electric Inc.) was then fixed to the sample using Kapton tape and UV curing resin, forming the electrochemical cell as shown in Extended Data Fig. 1**c**.

**Electrolyte solution**

The electrolyte solution was composed of ultrapure water (Ultrapure Water (H$_2$O), Fujifilm Wako Pure Chemical Co.), a chiral electrolyte ((1*S*)-(+)- or (1*R*)-(−)-camphor-10-sulfonic acid ((*S*)-CSA or (*R*)-CSA), Sigma-Aldrich), and an achiral supporting electrolyte (potassium chloride (KCl), Fujifilm Wako Pure Chemical Co.). These reagents were used directly as purchased. In all the experiments in the main text, CSA was used at a concentration of 0.25 mM, while KCl was maintained at 50 mM. The electrolyte solution in a volume of 30 μL, prepared in a nitrogen glove box (UNICO Ltd.), was added to the electrochemical cell. The cell was then sealed and removed from the glove box for the subsequent electrical measurements.

**Magnetoconductance (MC) measurement**

The electrical measurements were carried out utilizing a two-terminal method, employing a custom-made probe system. The system was integrated with a Keithley 2450 Source Measure Unit and an electromagnet to characterize the MC effect. All measurements were conducted under ambient conditions.

**First-principles calculation**

DFT calculations were performed with the VASP code[49] using the projected augmented wave method[50] and the Perdew-Burke-Ernzerhof exchange-correlation functional[51]. The plane-wave energy cutoff was set to 500 eV. A CSA molecule was placed in a cubic cell (edge length: 20 Å) with the molecular dipole moment aligned along the $z$-axis. The spin–orbit coupling was included in the calculation, so we used noncollinear spin calculations. The initial magnetization is given by adding 0.1 electrons, with the direction of the magnetic moment set to $+z$ or $-z$. Frozen phonon calculations were performed to obtain the vibrational modes.

We performed additional calculations to extract important vibrational modes. First, we calculated the electron–vibration coupling between electronic states and vibrational modes, the importance of which has been suggested in the literature[31]. We employed the SIESTA package[52] modified to compute electron–vibration coupling constants[53]. The electron–vibration interaction Hamiltonian is

$$\mathbf{H}_{\text{e-ph}} = \sum_m \sum_{i,j} M_{ij}^m c_i^\dagger c_j (b_m^\dagger + b_m), \tag{i}$$

$$M_{ij}^m = \sum_{A\nu} \left\langle i \left| \frac{\partial \mathbf{H}_e}{\partial R_{A\nu}} \right| j \right\rangle_{Q=0} Q_{A\nu}^m \sqrt{\frac{\hbar}{2M_A \omega_m}}, \tag{ii}$$

where $c_i^\dagger$ and $c_j$ are the creation and annihilation operators of an electron and $i$ is the index of the basis function. $b_m^\dagger$ and $b_m$ are those of vibration $m$, with frequency $\omega_m$ and eigenvector $Q_{A\nu}^m$, where $A$ and $\nu$ goes over atom and cartesian axes $x$, $y$, and $z$, respectively. $M_A$ is the mass of atom $A$. $R_{A\nu}$ is the displacement around the equilibrium position and $\mathbf{H}_e$ is the electron Hamiltonian. In order to calculate Eq. (ii) practically, we rewrite it as

$$\left\langle i \left| \frac{\partial \mathbf{H}_e}{\partial R_{A\nu}} \right| j \right\rangle = \frac{\langle i | \partial \mathbf{H}_e | j \rangle}{\partial R_{A\nu}} - \langle i' | \mathbf{H}_e | j \rangle - \langle i | \mathbf{H}_e | j' \rangle, \tag{iii}$$

where $|i'\rangle = \partial |i\rangle / \partial R_{A\nu}$ is the change in the basis function due to the atomic displacement. This term appears because SIESTA uses a linear combination of atomic orbitals. Using the completeness condition via overlap integral $\mathbf{S}$,

$$\sum_{ij} |i\rangle (\mathbf{S}^{-1})_{ij} \langle j| = 1,$$

we obtain

$$\left\langle i \left| \frac{\partial \mathbf{H}_e}{\partial R_{A\nu}} \right| j \right\rangle = \frac{\langle i | \partial \mathbf{H}_e | j \rangle}{\partial R_{A\nu}} - \sum_{kl} \langle i' | k \rangle (\mathbf{S}^{-1})_{kl} \langle l' | \mathbf{H}_e | j \rangle - \sum_{kl} \langle i | \mathbf{H}_e | k \rangle (\mathbf{S}^{-1})_{kl} \langle l | j \rangle. \tag{iv}$$

This can be calculated by slightly modifying the SIESTA program to output the Hamiltonian and the overlap matrix elements in each coordinate. Note that only the linear coupling term is assumed in Eq. (i), therefore the Born approximation, where the multiple scattering interactions are excluded, is already assumed. For the later analysis, it is useful to transform the atomic basis to the molecular orbital basis. In terms of molecular orbital basis $I$ and $J$, Eq. (ii) can be rewritten as

$$M_{IJ}^m = \sum_{A\nu} \left( \left\langle I \left| \frac{\partial \mathbf{H}_e}{\partial R_{A\nu}} \right| J \right\rangle - \varepsilon_J \left\langle \frac{\partial}{\partial R_{A\nu}} I \middle| J \right\rangle - \varepsilon_I \left\langle I \middle| \frac{\partial}{\partial R_{A\nu}} J \right\rangle \right) Q_{A\nu}^m \sqrt{\frac{\hbar}{2M_A \omega_m}}, \tag{v}$$

where $\varepsilon_I$ is the energy of the molecular orbital $I$. When $I$ is equal to $J$, the coupling is related to the deformation of the molecular orbital according to the normal mode vector

$$M_{II}^m = \sum_{A\nu} \frac{\partial \varepsilon_I}{\partial R_{A\nu}} Q_{A\nu}^m \sqrt{\frac{\hbar}{2M_A \omega_m}}. \tag{vi}$$

Here, we calculated the electron–vibration coupling constant for the highest occupied molecular orbital (HOMO) and the lowest unoccupied molecular orbital (LUMO) as

$$\lambda^m = \lambda^m_{\text{HOMO}} + \lambda^m_{\text{LUMO}}, \tag{vii}$$

where $\lambda^m_I = (M^m_{II})^2/\omega_m$. For the SIESTA calculations, we used the PBE functional and the double-$\zeta$ plus polarization basis set. The cutoff for the real space grid was set to 300 Ry.

Second, we calculated the vibrational circular dichroism (VCD) spectrum using the Gaussian 16 program[54] at the B3LYP/6-31G (d,p) level of theory to investigate the coupling between each vibrational mode and the molecular chirality. Circular dichroism (CD) is the differential absorption of left and right circularly polarized light.

$$\Delta a = a^L + a^R. \tag{viii}$$

CD can be expressed in terms of rotational strength as[55]

$$\Delta a \propto R = \omega^{-1} \frac{\partial \mu}{\partial t} \otimes m, \tag{iv}$$

where $\mu$ and $m$ are the electric dipole magnetic dipole moments, respectively. Therefore, a large VCD indicates a large interaction between the molecular vibration and the magnetic moment.

Finally, we calculated the spin–vibration coupling, which is the change in the magnetization according to the displacement along the normal mode (~0.1 Å) using the VASP package.

All the results are shown in Extended Data Table 1. We adopted the frequency calculated with VASP, because the order of vibrational modes was the same between the three packages, although the frequencies were slightly different. Although calculation results show that every mode has finite spin–vibration coupling and VCD, we selected a representative vibrational mode as shown in Extended Data Fig. 3, that is, C=O stretching mode (1770 cm$^{-1}$), which has the largest electron–vibration coupling constant.

**Acknowledgements**

We thank Sachiko Kamisaka and Yukiko Kato of The University of Tokyo for their assistance. We also thank Kouta Kondou of RIKEN, Daigo Miyajima of The Chinese University of Hong Kong, Masayuki Suda of Kyoto University, Shun Watanabe of The University of Tokyo for discussion, and Tomoya Higo and Satoru Nakatsuji of The University of Tokyo for magnetooptical Kerr effect measurements. This work was partially supported by JSPS-KAKENHI (Nos. JP21H01032, JP22K18320, JP22H00290, JP22H04964, JP23H00091, JP23H00291, and JP24H02234), JST-ASPIRE (No. JPMJAP2317), Spintronics Research Network of Japan (Spin-RNJ), and MEXT Initiative to Establish Next-Generation Novel Integrated Circuit Centers (X-NICS) (No. JPJ011438).

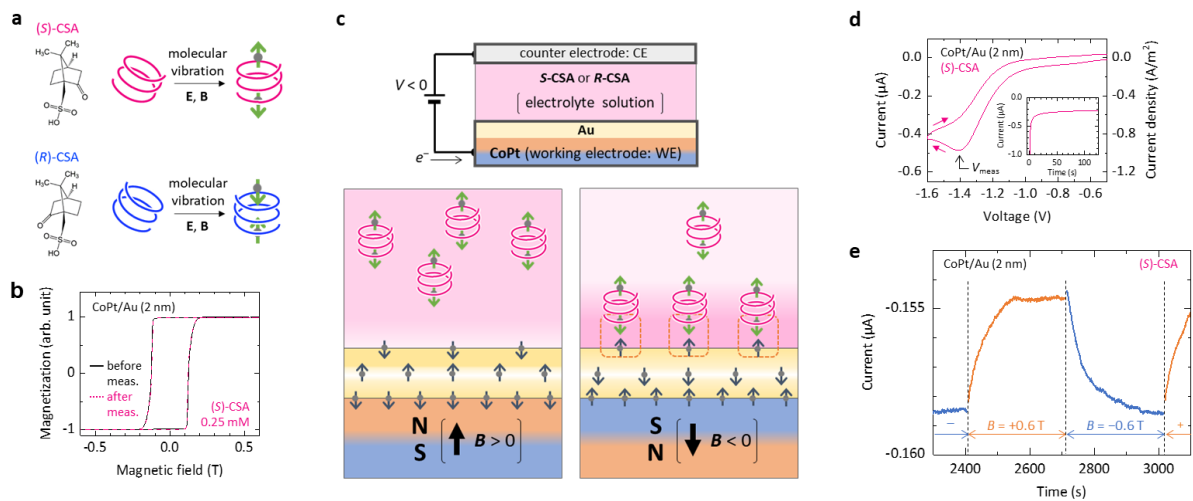

**Fig. 1 | Magnetic interactions driven by molecular vibration at interfaces and experimental design. a,** Schematic of spin polarization driven by molecular vibration under electric (**E**) and magnetic (**B**) fields in chiral molecules, utilizing (1*S*)-(+)- camphor-10-sulfonic acid ((*S*)-CSA) and (1*R*)-(−)-camphor-10-sulfonic acid ((*R*)-CSA) as chiral molecules. **b,** Magnetization hysteresis curves for the CoPt electrode measured by the magnetooptical Kerr effect, observed both before and after electrical measurements in an electrochemical cell, with the magnetic field oriented perpendicular to the film plane. **c,** A schematic illustration of the experimental setup. **d,** Typical current–voltage characteristics of the electrochemical cell containing (*S*)-CSA at 0.25 mM and KCl at 50 mM, with a voltage sweep rate of 60 mV/s. The inset shows chronoamperometry results under a steady voltage ($V_{\text{meas}}$). **e,** Representative results of magnetoconductance measurements. The working electrode had a surface area of 0.5 mm$^2$. Gray and green arrows in the figure represent the spin angular momentum (**S**) and its time derivative (d**S**/d*t*), respectively.

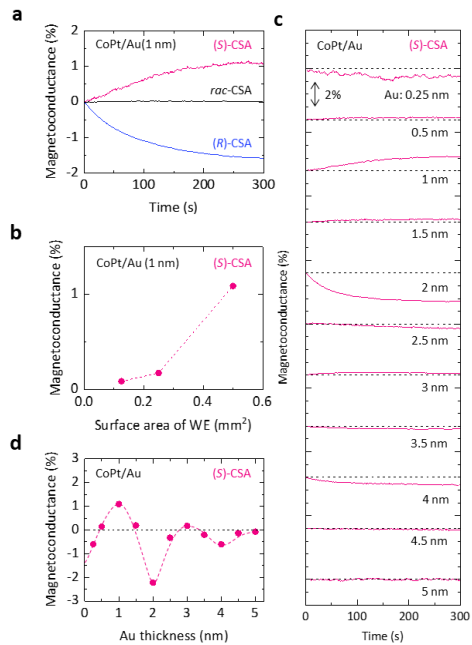

**Fig. 2 | Time-resolved magnetoconductance (MC) measurements. a,** Chirality dependence on the MC effect using a CoPt/Au (1 nm) electrode. **b,** Variation of MC effect with different surface areas of the working electrode (WE). **c,** Influence of Au thickness on the MC effect. **d,** The MC ratio as a function of the thickness of the Au spacer. In all data shown in Fig. 2, the concentrations of CSA and KCl are set at 0.25 mM and 50 mM, respectively. The dashed curve represents the guide to the eye. Except for Fig. 2**b**, the surface area of the working electrode is fixed at 0.5 mm$^2$.

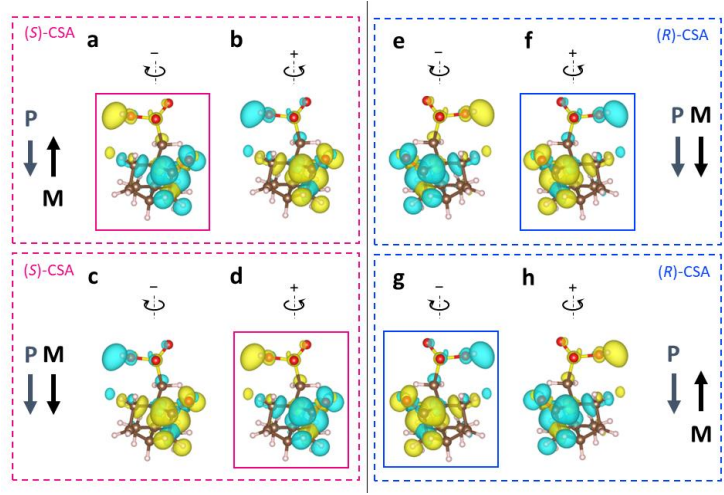

**Fig. 3 | First principles study. a, b,** A schematic diagrams of the computational model for (*S*)-CSA showing spin density under antiparallel electric polarization (**P**) and magnetization (**M**) configurations. The contours indicate changes in spin angular momentum (d**S**/d*t*) for positive (+) and negative (−) displacements; yellow indicates d**S**/d*t* > 0 (upward) and cyan indicates d**S**/d*t* < 0 (downward). This model identifies a key molecular vibrational mode at 1770 cm$^{-1}$. **c, d,** Identical calculations for (*S*)-CSA, but with the parallel alignment of **P** and **M**. **e-h,** Similar analyses for (*R*)-CSA. Considering the vibrational angular momentum and spin–vibration coupling in a magnetic field, molecular vibrations can cause variation in the duration of polarities where d**S**/d*t* > 0 and d**S**/d*t* < 0, leading to a predominance of d**S**/d*t* density in Figs. 3**a**, 3**d**, 3**f** and 3**g**, as marked by solid squares.

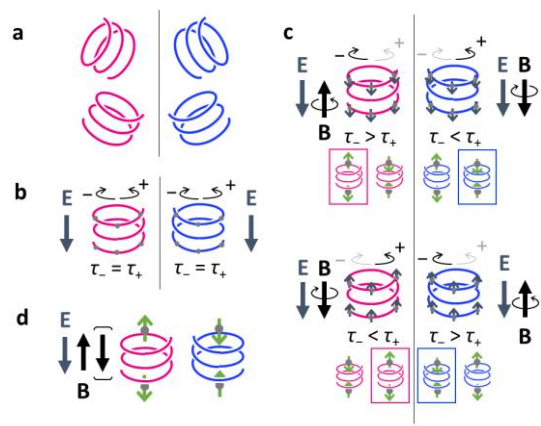

**Fig. 4 | Spin polarization driven by molecular vibration. a, b,** Chiral molecules align under an electric field (**E**), orienting their molecular structures. **c,** A magnetic field (**B**) results in a chirality-independent spin polarization. At finite temperatures, the interplay between electron–vibration coupling and spin–orbit interaction can result in chirality-dependent spin polarizations at the edge of the chiral molecules. The influence of spin–vibration coupling under a magnetic field is expected to cause variations in the durations of half-periods $\tau_+$ and $\tau_-$, due to vibrational angular momentum. Gray and green arrows represent the spin angular momentum (**S**) and its time derivative (d**S**/d$t$), respectively. **d,** The time derivative of the spin angular momentum (d**S**/d$t$) in right-handed and left-handed molecules, affected by both **E** and **B**. The direction of **B** affects the spin polarization polarity and the relative lengths of the half-periods. As a result, the dominant d**S**/d$t$ polarity remains consistent regardless of the direction of **B**, being determined entirely by the molecular handedness.